\documentclass[twocolumn,showpacs]{revtex4}
\usepackage{graphicx}
\usepackage{epstopdf}
\usepackage{bm}

\begin{document}
\title{Observation of photon-assisted tunneling in optical lattices}

\author{C. Sias, H. Lignier, Y. P. Singh, A. Zenesini, D. Ciampini, O. Morsch and E. Arimondo}
\affiliation{CNR-INFM, Dipartimento di Fisica `E. Fermi', Largo
Pontecorvo 3, 56127 Pisa, Italy}

\begin{abstract}
We have observed tunneling suppression and photon-assisted
tunneling of Bose-Einstein condensates in an optical lattice
subjected to a constant force plus a sinusoidal shaking. For a
sufficiently large constant force, the ground energy levels of the
lattice are shifted out of resonance and tunneling is suppressed;
when the shaking is switched on, the levels are coupled by
low-frequency photons and tunneling resumes. Our results agree
well with theoretical predictions and demonstrate the usefulness
of optical lattices for studying solid-state phenomena.
\end{abstract}


\maketitle

A number of experiments in recent years have shown that
Bose-Einstein condensates (BECs)~\cite{morsch_review} loaded into
optical lattices are well suited to simulating solid state
systems~\cite{bloch_natphys,bloch_review}. Optical lattices are
created by crossing two or more laser beams, and the resulting
periodic potential landscapes (arising from the ac-Stark shift
exerted on the condensate atoms) are intrinsically defect-free,
their lattice wells have controllable depths, and it is possible
to move or accelerate the entire structure. This flexibility has
made it possible to study dynamical effects such as Bloch
oscillations~\cite{morsch01} and resonant tunneling~\cite{sias07}
as well as ground-state quantum properties such as the
Mott-insulator transition~\cite{greiner02a}. More recently, the
coherent suppression of inter-well tunneling by strong driving of
the lattice has been demonstrated~\cite{lignier07}. In this
Letter, we explore an effect~\cite{eckardt05} that is analogous to
photon-assisted tunneling in solids and arises from the interplay
between static acceleration and strong driving of the lattice. We
observe two regimes, a linear and a nonlinear one, with different
dependencies of the observed tunneling on the theoretically
predicted behaviour.

\begin{figure}[ht]
\includegraphics[width=8cm]{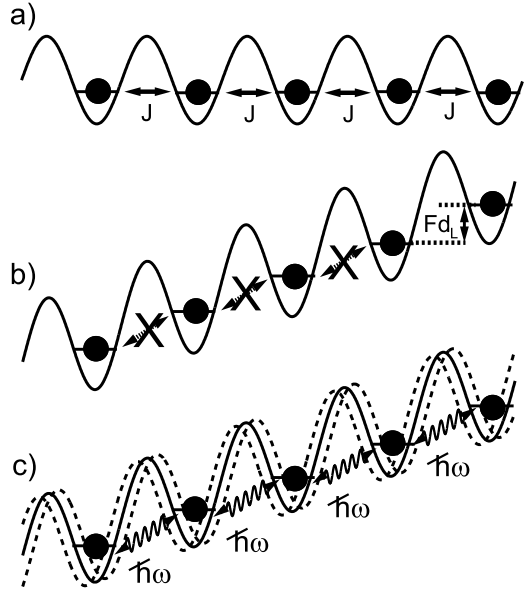}
\caption{\label{figure1} Photon-assisted tunneling. (a) In an
optical lattice at rest, the ground-state levels are resonantly
coupled, leading to a tunneling energy $J$. (b) When a linear
potential is applied, e.g. by accelerating the lattice, the levels
are shifted out of resonance and tunneling is suppressed. (c) If
the lattice is now periodically shaken at an appropriate
frequency, the levels can again be coupled through photons of
energy $\hbar \omega$ and tunneling is partially restored.}
\end{figure}

Photon-assisted tunneling occurs when adjacent potential wells
whose ground states are tuned out of resonance by a static
potential are coupled by photons (see Fig. 1). The static force
leads to a suppression of resonant tunneling between the ground
states. This suppression and the related Wannier-Stark
localization of the wavefunction have been intensively discussed
in the theoretical literature~\cite{rossi98,gluck02}. In this
work, we report a direct measurement of this suppression based on
the spatial tunneling of the condensate atoms. When photons of an
appropriate frequency are present whose energy bridges the gap
created by the static potential, tunneling is (partly) restored.
In solid state systems, the photons are typically in the microwave
frequency range and the static potential is provided by an
electric bias field applied to the structure. So far,
photon-assisted tunneling has been observed in superconducting
diodes~\cite{tien63}, semiconductor
superlattices~\cite{guimaraes93,keay95,keay95b} and quantum
dots~\cite{kouwenhoven94,oosterkamp97}.

Our system consists of a BEC inside a one-dimensional optical
lattice. The static potential is provided by a constant
acceleration of the lattice, resulting in a constant force $F$ in
the lattice rest frame and hence in a potential difference $\Delta
E=Fd_L$ between adjacent wells a distance $d_L$ apart. The role of
the photons is played by a periodic shaking of the lattice at
frequency $\omega$ that leads to the creation of sidebands around
the carrier frequency of the laser beam. In the limit of
sufficiently deep lattice wells and neglecting higher-lying energy
levels, our system can be described by the
Hamiltonian~\cite{eckardt05}
\begin{eqnarray}
\hat{H}_0 = &-&J\sum_{\langle
i,j\rangle}(\hat{c}_i^\dagger\hat{c}_j+\hat{c}_j^\dagger\hat{c}_i)+\frac{U}{2}\sum_j
\hat{n}_j(\hat{n}_j-1)+ \\ &+& \Delta E\sum_j
j\hat{n}_j+K\cos(\omega t)\sum_j j\hat{n}_j, \nonumber
\end{eqnarray}
where $\hat{c}_i^{(\dagger)}$ are the boson creation and
annihilation operators on site $i$,
$\hat{n}_i=\hat{c}_i^{\dagger}\hat{c}_i$ are the number operators,
and $K$ and $\omega$ are the strength and angular frequency of the
shaking, respectively. The first line of this equation is the
Bose-Hubbard model~\cite{jaksch_98} with the tunneling matrix
element $J$ and the on-site interaction term $U$ (in a BEC, the
on-site interaction is due to atom-atom collisions and hence
proportional to the $s$-wave scattering length and the density of
the BEC). In the second line, the first term describes the
constant potential, whereas the second term represents the
sinusoidal shaking of the lattice. While for a sufficiently strong
linear potential inter-well tunneling is suppressed, leading to
Wannier-Stark localization, recent theoretical
work~\cite{eckardt05} predicts that the shaking term can partially
restore it, leading to an effective tunneling rate
\begin{equation}\label{eq2}
|J_\mathrm{eff}(K_0)/J|=|\mathcal{J}_{n}(K_0)|
\end{equation}
when the resonance condition
\begin{equation}\label{eq3}
n\hbar\omega=Fd_L
\end{equation}
is satisfied, where $n$ is an integer denoting the order of the
photon-assisted resonance, $\mathcal{J}_{n}$ is the $n$-th order
ordinary Bessel function, and $K_0=K/\hbar \omega$ is the
dimensionless parameter characterizing the shaking amplitude.

In our experiment we produced BECs of $^{87}\mathrm{Rb}$
containing around $5\times 10^4$ atoms in a crossed optical dipole
trap. The two dipole traps were created by gaussian laser beams at
$1030\,\mathrm{nm}$ wavelength and a power of around
$1\,\mathrm{W}$ per beam focused to waists of $50\,\mathrm{\mu
m}$, and the frequencies of the resulting trapping potentials
could be controlled independently. Subsequently, the BECs held in
the dipole trap were loaded into an optical lattice created by two
counter-propagating gaussian laser beams ($\lambda =
852\,\mathrm{nm}$) with $120\,\mathrm{\mu m}$ waists by ramping up
the power of the lattice beams in about $50\,\mathrm{ms}$. The
resulting periodic potential $V(x)=V_0\sin^2(\pi x/d_L)$ had a
lattice spacing $d_L=\lambda/2=426\,\mathrm{nm}$ and its depth
$V_0$ was measured in units of the recoil energy $E_{\rm rec}=
\hbar^2 \pi^2 / (2m d_L^2)$, where $m$ is the mass of the Rb
atoms. By introducing a frequency difference $\Delta \nu$ between
the two lattice beams (using acousto-optic modulators, which also
control the power of the beams), the optical lattice could be
moved at a velocity $v=d_L\Delta \nu$ or accelerated with an
acceleration $a=d_L\frac{d\Delta\nu}{dt}$. In order to
periodically shake the lattice, $\Delta \nu$ was sinusoidally
varied with frequency $\omega$ and amplitude
$\Delta\nu_{\mathrm{max}}$ leading to a time-varying force (in the
rest frame of the lattice)
\begin{equation}
F(t)= m\omega d_L \Delta\nu_{\mathrm{max}}\cos(\omega t)=
F_{\mathrm{max}}\cos(\omega t).
\end{equation}
The dimensionless shaking parameter $K_0$ is then given by
\begin{equation}
K_0=K/\hbar\omega=md_L^2\Delta\nu_{\mathrm{max}}/\hbar=\pi^2\Delta\nu_{\mathrm{max}}/2\omega_{\mathrm{rec}}.
\end{equation}

Our method for measuring the effective tunneling parameter
$|J_\mathrm{eff}/J|$ (where the modulus indicates that this
measurement is not sensitive to the sign of $J_\mathrm{eff}$) is
based on the free expansion of the BEC~\cite{anker05,lignier07}
confined only radially but free to move along the direction of the
lattice (the dipole trap frequency in that direction being on the
order of a few Hz and hence negligible for our purposes). After
condensation was reached in the crossed dipole trap and the
optical lattice had been ramped up, the trapping beam
perpendicular to the lattice direction was suddenly switched off.
Subsequently, the in-situ width of the BEC in the lattice
direction was measured by flashing on a resonant beam and imaging
the shadow cast by the BEC on a CCD camera. In the experiments
with an accelerated and / or shaken lattice, $|J_\mathrm{eff}/J|$
was determined by measuring the expansion for the same lattice
depth both in the driven case and without the driving.

\begin{figure}[ht]
\includegraphics[width=8cm]{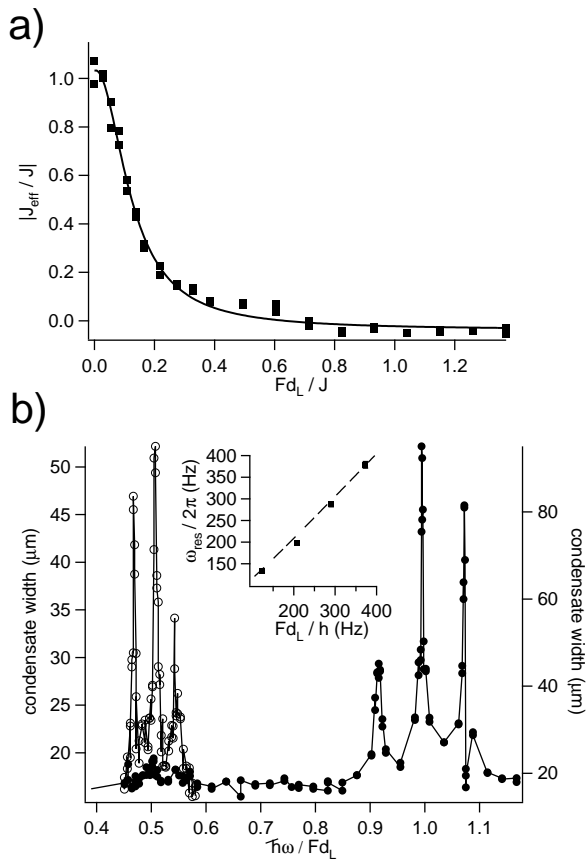}
\caption{\label{figure2} (a) Suppression of tunneling by a linear
potential. Shown here is the normalized effective tunneling
parameter $|J_\mathrm{eff}/J|$ as a function of the linear
potential $Fd_L$ in units of the tunneling energy. When
$Fd_L/J\approx 1$, the ground state levels are shifted out of
resonance and tunneling is suppressed almost completely. The solid
line is a Lorentzian fit with a half-maximum half-width of $0.13$.
(b) Photon-assisted tunneling resonances in a shaken lattice. For
a fixed linear potential $Fd_L/h=380\,\mathrm{Hz}$, the condensate
width after $400\,\mathrm{ms}$ of free expansion is plotted as a
function of the normalized shaking frequency $\hbar\omega/Fd_L$.
The fixed shaking parameter was $K_0=1.8$ for the one-photon
resonance (solid circles) and $K_0=3.1$ for the two-photon
resonance (open circles), corresponding to the first maximum of
the $\mathcal J_1$ and $\mathcal J_2$ Bessel functions,
respectively. For both graphs, $V_0/E_\mathrm{rec}=5$ and
$J/h=380\,\mathrm{Hz}$.}
\end{figure}

In a preliminary experiment, we studied the tunneling suppression
caused by shifting adjacent ground states out of resonance through
a constant force $F$ acting on the BEC inside a lattice that was
subjected to an acceleration $a$. Typical accelerations in our
experiment were between $0$ and $4\,\mathrm{ms^{-2}}$, meaning
that for an expansion time of $400\,\mathrm{ms}$ the lattice (and
therefore the BEC) would have been displaced by up to
$32\,\mathrm{cm}$, two orders of magnitude more than the
field-of-view of our imaging system. Also, a displacement of more
than $100\,\mathrm{\mu m}$ along the lattice direction would have
led to a restoring force of the longitudinal harmonic trap created
by the radial dipole trap beam corresponding to an acceleration
$a_\mathrm{restore}>0.1\,\mathrm{ms^{-2}}$. Therefore, in order to
be able to achieve a high resolution in our measurements of the
expansion rate (and hence $J$) of the condensate, implying a long
expansion time, whilst keeping the displacement of the lattice
below $\approx 100\,\mathrm{\mu m}$, we used a rectangular
acceleration profile that alternated between $+a$ and $-a$ and
therefore `rocked' the lattice back and forth. In this way, the
modulus of the resulting force and hence the energy shift between
adjacent wells was constant, while the mean position of the
lattice (and hence the BEC) remained close to the center of the
dipole trap. In order to separate the frequency regimes of the
rocking and the shaking motion, we chose a `rocking frequency' of
$30\,\mathrm{Hz}$, which was much smaller than the resonant
frequencies for photon-assisted tunneling (typically around
$150-400\,\mathrm{Hz}$).

Figure 2 (a) shows the results of our measurements of
$|J_\mathrm{eff}|$ in an accelerated (rocked) lattice. As
expected, when the energy difference $Fd_L$ between adjacent
levels is increased, resonant tunneling is reduced and, for
$Fd_L\approx J$, completely suppressed (as recently also observed
for single-atom tunneling in a double-well
structure~\cite{folling07}). In this limit, the energy levels in
the individual wells can be viewed as Wannier-Stark levels. Our
data are fitted very well by a Lorentzian, but to our knowledge
there is no analytical prediction for such a dependence in the
theoretical literature. In~\cite{gluck02}, an expression for the
wavefunction in a tilted lattice as a function of the applied
force is given, but an analytical calculation of
$J_\mathrm{eff}(F)$ has yet to be done.

Tunneling between the on-site levels shifted out of resonance by
the static acceleration can be partially restored by sinusoidally
shaking the lattice at a frequency $\omega$ satisfying the
resonance condition of Eq.~\ref{eq3}. The shaking of the lattice
effectively creates low-frequency `photons' that bridge the energy
gap between adjacent wells with $n$ such photons (see Fig. 1 (c)).
Figure 2 (b) shows the condensate width after $400\,\mathrm{ms}$
of free expansion inside a rocked lattice with $Fd_L/J=1$ as a
function of the shaking frequency $\omega$. One clearly sees two
photon-assisted tunneling peaks at $\hbar\omega=Fd_L$ (for
$K_0=1.8$, where ${\mathcal J}_1(K_0)$ has its first maximum) and
at $2\hbar\omega=Fd_L$ (for $K_0=3.1$, the first maximum of
${\mathcal J}_2(K_0)$). The peaks are extremely narrow, with a
width of $\approx 3\,\mathrm{Hz}$ (when we repeated the experiment
with smaller values of $Fd_L$, this width increased slightly). The
two side-peaks of each peak are evidence of additional
photon-assisted tunneling events due to the rocking motion of the
lattice (for the two-photon resonance at $2\hbar\omega=Fd_L$, the
side-peaks are at half the distance to the main peak compared to
the one-photon case, as expected).

\begin{figure}[ht]
\includegraphics[width=8cm]{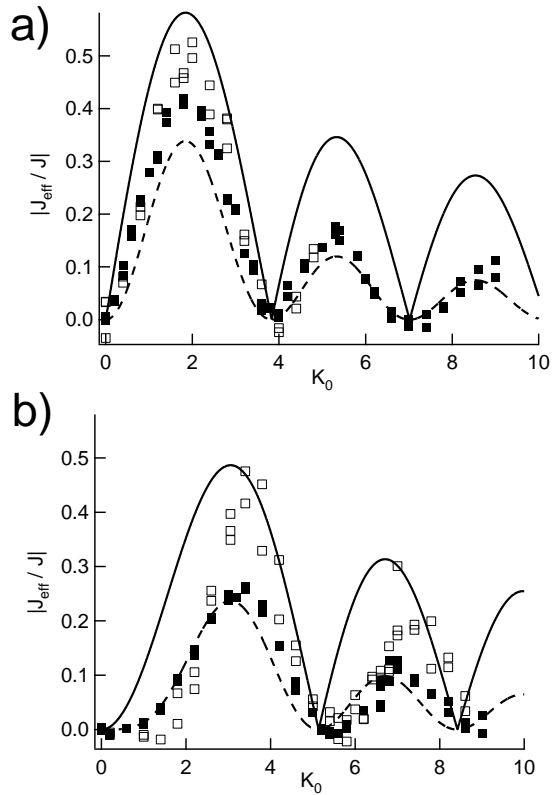}
\caption{\label{figure3} Photon-assisted tunneling as a function
of the shaking parameter $K_0$. Shown here are the one-photon
resonance at $\omega/2\pi=380\,\mathrm{Hz}$ (a) and the two-photon
resonance at $\omega/2\pi=190\,\mathrm{Hz}$ (b). In both graphs,
the full and open squares are the measurements for $N\approx
5\times 10^4$ and $N\approx 0.5\times 10^4$, respectively. The
solid lines are the moduli of the $\mathcal J_{1}(K_0)$ and
$\mathcal J_{2}(K_0)$ Bessel functions, respectively, whereas the
dashed lines are the squares of these functions. The lattice depth
$V_0/E_\mathrm{rec}=5$ and the constant force
$Fd_L/h=380\,\mathrm{Hz}$ in this experiment, with a free
expansion time $t=100\,\mathrm{ms}$.}
\end{figure}

Finally, we studied the dependence of the effective tunneling rate
on the shaking parameter $K_0$. Figure 3 summarizes our results
for the one-photon and two-photon resonances. Theory
predicts~\cite{eckardt05} that the effective tunneling rates
$|J_\mathrm{eff}/J|$ for resonances of $n$-th order should vary as
an $n$-th order ordinary Bessel function (see Eq.~\ref{eq2}).
While qualitative agreement between experiment and theory is good,
with the positions of the maxima and minima of the one- and
two-photon resonances as a function of $K_0$ coinciding perfectly
with theoretical predictions, the absolute values of
$|J_\mathrm{eff}/J|$ lie consistently below the theoretical curves
by a factor of about $1.3$.

Interestingly, quantitative agreement between experiment and
theory is better if we use the squares of the Bessel functions
rather than their moduli. A dependence of the photon-assisted
tunneling rate on the square of the Bessel function is expected,
e.g., for experiments on Josephson junctions irradiated by
microwaves and, more generally, if the tunneling in a multi-well
structure is sequential rather than coherent. The difference
between coherent and sequential tunneling has been extensively
studied in the theoretical literature~\cite{buttiker88}.
Sequential tunneling would require a dephasing mechanism between
two successive tunneling events. In our experiments, such a
mechanism could be the dynamical instability inside the optical
lattice~\cite{cristiani04,fallani04,mun07}: owing to the constant
applied acceleration $a$, the BEC moves through the Brillouin zone
and hence through the dynamically unstable region within the
Brillouin zone between about $0.5\,p_\mathrm{rec}$ (where
$p_\mathrm{rec}=\pi\hbar/d_L$ is the recoil momentum ) and the
zone edge. For the values of the intrinsic nonlinearity of the BEC
in our experiments (determined by the number of atoms and the trap
frequencies), we found in previous experiments~\cite{cristiani04}
(and have verified for this work) that the BEC loses its phase
coherence after a few milliseconds inside the unstable region,
which is comparable to the Bloch periods in the present
experiment. Since the effective tunneling frequencies
$J_\mathrm{eff}$ in our experiment are less than
$100\,\mathrm{Hz}$, the corresponding dephasing rate is almost an
order of magnitude larger and hence it is likely that dephasing of
neighbouring wells occurs between two tunneling events.

In future experiments one might use, e.g., Feshbach resonances in
order to tune the nonlinearity and hence move from the strongly
nonlinear regime to the linear regime in order to test our
hypothesis. As a preliminary test, we have repeated the
measurement of the one-photon assisted tunneling rate as a
function of $K_0$ with the smallest atom number that allowed us to
measure the free expansion ($N\approx 0.5\times 10^4$), resulting
in a condensate density that was about a factor of 3 smaller.
Again, we obtained qualitatively similar results to the
measurements with $N\approx 5\times 10^4$, but this time the
absolute values of $|J_\mathrm{eff}/J|$ agreed better with the
linear Bessel-function prediction (see Fig. 3).

A further indication that dephasing might be responsible for the
observed deviation from the linear Bessel functions comes from our
measurements of the dynamical suppression of tunneling in a shaken
lattice \textit{without} linear acceleration~\cite{lignier07}, in
which our measured values for $J_\mathrm{eff}/J$ agreed perfectly
with the linear Bessel function prediction. In that system, the
BEC does not cross the unstable region of the Brillouin zone, and
we have experimentally verified that during the shaking the BEC
retains its phase coherence.

If the two regimes of photon-assisted tunneling observed in our
experiment do, indeed, correspond to coherent and sequential
tunneling, our system is ideally suited to studying the cross-over
between these two extremes in a well-controlled way. However, we
also have to consider the possibility that other effects play a
role. In particular, it is conceivable that a self-trapping
mechanism is present that depends on the relative magnitude of the
nonlinearity $U$ and the effective tunneling parameter
$J_\mathrm{eff}$. Self-trapping in static optical lattices has
already been observed~\cite{anker05} and could, if present also in
our strongly driven system, lead to a suppression of the tunneling
rate compatible with our observations, also for the case of a
static force as in Fig. 2(a).  On the other hand, our system is
only in a self-trapping regime (as calculated from
$U/h\approx10-30$ Hz and $J_{\mathrm{eff}}$) in a small region
around the zeroes of the Bessel functions, whereas we observe the
largest deviation from the linear prediction close to the local
maxima.

In summary, we have demonstrated photon-assisted tunneling of BECs
in a linearly accelerated and sinusoidally shaken optical lattice.
Our results agree quantitatively with recent theoretical
predictions and show the need for a theoretical investigation into
the difference between the (roughly) linear and the nonlinear
regimes for which we observe difference dependencies of the
effective tunneling rate on the predicted linear Besse-function
scaling.

This work was supported by OLAQUI and MIUR-PRIN. The authors would
like to thank M. Holthaus for illuminating discussions.

\bibliographystyle{apsrmp}

\begin{thebibliography}{99}

\bibitem{morsch_review}
O. Morsch and M. Oberthaler, Rev. Mod. Phys. \textbf{78}, 179
(2006).

\bibitem{bloch_natphys}
I. Bloch, Nat. Phys. \textbf{1}, 23 (2005).

\bibitem{bloch_review}
I. Bloch, J. Dalibard, and W. Zwerger, arXiv:0704.3011 (2007).

\bibitem{morsch01}
O. Morsch, J.H. M\"uller, M. Cristiani, D. Ciampini, and E.
Arimondo, Phys. Rev. Lett. \textbf{87}, 140402 (2001).

\bibitem{sias07}
C. Sias \textit{et al.}, Phys. Rev. Lett. \textbf{98}, 120403
(2007).

\bibitem{greiner02a}
M. Greiner, O. Mandel, T. Esslinger, T.~W. H\"ansch, and I. Bloch,
Nature \textbf{415}, 39 (2002).

\bibitem{lignier07}
H. Lignier \textit{et al.}, arXiv:0707.0403 (2007).

\bibitem{eckardt05}
A. Eckardt, T. Jinasundera, C. Weiss, and M. Holthaus, Phys. Rev.
Lett. \textbf{95}, 200401 (2005).

\bibitem{rossi98} F. Rossi, Semicond. Sci. Technol. \textbf{13},
147 (1998).

\bibitem{gluck02} M. Gl\"{u}ck, A. R. Kolovsky, and H. J. Korsch,
Physics Reports \textbf{366}, 103 (2002).

\bibitem{tien63}
P.K. Tien and J.P. Gordon, Phys. Rev. \textbf{129}, 647 (1963).

\bibitem{guimaraes93}
P.S.S. Guimar\~{a}es {\it et al.}, Phys. Rev. Lett. \textbf{70},
3792 (1993).

\bibitem{keay95}
B.J. Keay {\it et al.}, Phys. Rev. Lett. \textbf{75}, 4098 (1995).

\bibitem{keay95b}
B.J. Keay {\it et al.}, Phys. Rev. Lett. \textbf{75}, 4102 (1995).


\bibitem{kouwenhoven94}
L.P. Kouwenhoven {\it et al.}, Phys. Rev. Lett. \textbf{73}, 3443
(1994).

\bibitem{oosterkamp97}
T.H. Oosterkamp, L.P. Kouwenhoven, A.E.A. Koolen, N.C. van der
Vaart, and C.J.P.M. Harmans, Phys. Rev. Lett. \textbf{78}, 1536
(1997).

\bibitem{jaksch_98} D. Jaksch, C. Bruder, J.~I. Cirac, C.~W. Gardiner, and P. Zoller,
Phys. Rev. Lett. \textbf{81}, 3108 (1998).

\bibitem{anker05}
Th. Anker \textit{et al.}, Phys. Rev. Lett. \textbf{94}, 0204003
(2005).

\bibitem{folling07} S. F\"{o}lling \textit{et al.}, Nature
\textbf{448}, 1029 (2007).

\bibitem{buttiker88} M. B\"{u}ttiker, IBM J. Res. Develop.
\textbf{32}, 63 (1988).

\bibitem{cristiani04} M. Cristiani, \textit{et al.}  Opt.
Express {\bf 12}, 4 (2004).

\bibitem{fallani04} L. Fallani, \textit{et al.},  Phys. Rev. Lett. {\bf 93}, 140406 (2004).

\bibitem{mun07} J. Mun,  \textit{et al.}, arXiv:0706.3946 (2007).

\end{thebibliography}

\end{document}